\pdfoutput=1
\documentclass[preprint,pre,showpacs,preprintnumbers,amsmath,amssymb,longbibliography]{revtex4-1}

\usepackage{graphicx}% Include figure files
\usepackage{dcolumn}% Align table columns on decimal point
\usepackage{bm}% bold math
\usepackage{epstopdf}% pdflatex command active even for eps figures
\usepackage{hyperref}% Creates hyperlinks

	\usepackage{fancyheadings}
\renewcommand{\thispagestyle}[1]{} % do nothing

\begin{document}

\title{The Pair Approximation method for the ferromagnetic Heisenberg model with spin $S=1$ and arbitrary range of interactions. Application for the magnetic semiconductor CrIAs}
% Force line breaks with \\

\author{T. Balcerzak}
\email{tadeusz.balcerzak@uni.lodz.pl}
\homepage[]{https://orcid.org/0000-0001-7267-992X}
\author{K. Sza{\l}owski}
\email{karol.szalowski@uni.lodz.pl}
\homepage[]{https://orcid.org/0000-0002-3204-1849}
\affiliation{%
University of \L\'{o}d\'{z}, Faculty of Physics and Applied Informatics, Department of Solid State Physics,\\
ulica Pomorska 149/153, 90-236 \L\'{o}d\'{z}, Poland
}%

\date{\today}% It is always \today, today,
             %  but any date may be explicitly specified

\begin{abstract}

The Pair Approximation method has been formulated for the isotropic ferromagnetic Heisenberg model with spin $S=1$. The exchange interactions of arbitrary range have been taken into account. The single-ion anisotropy has been considered as well as the external magnetic field. Within the method, the Gibbs free-energy has been derived, from which all thermodynamic properties can be self-consistently obtained. In order to illustrate the developed formalism, the numerical calculations have been performed for CrIAs planar magnetic semiconductor, a hypothetical material whose existence has been recently predicted by the Density Functional Theory-based calculations. For this model material, all the relevant thermodynamic magnetic properties have been studied. The numerical results have been presented in the figures and discussed.

\end{abstract}

%\pacs{ }% PACS, the Physics and Astronomy Classification Scheme.

\keywords{Heisenberg model \sep Pair Approximation \sep spin $S=1$ \sep Gibbs energy  \sep magnetic properties \sep 2D magnetic semiconductor}
\maketitle

%section 1
\section{\label{intro}Introduction}

The Pair Approximation (PA) method originates from a more general Cluster Variational Method (CVM) formulated by Kikuchi \cite{Kikuchi1951}. The CVM represents a systematic cluster approach in which the 1st approximation, based on single-atom clusters, corresponds to Molecular Field Approximation (MFA), while the 2nd approximation, i.e., PA, based on two-atom clusters, is equivalent to Bethe approximation \cite{Bethe1935,Pelizzola2005}. It has also been shown \cite{Strieb1963, Morita1966} that PA is, to some extent, analogous to the Constant Coupling Approximation of Kasteleijn and Kranendonk \cite{Kasteleijn1956a}. 
Higher approximations within CVM, using square and cubic clusters, have also been introduced by Kikuchi \cite{Kikuchi1951} in a systematic way.

In the progress of further studies, the CVM, and at the same time PA, have been mastered with a tendency to clarify and simplify the formalism \cite{Morita1957, Morita1966, Strieb1963, Morita1972, Barry1975, Kinel1976, Tucker1987, An1988, Morita1989, Ma1990, Bukman1991, Rosengren1993, Katsura1996}.
This enabled application of the method to many physical problems, especially in the field of magnetism \cite{Balcerzak1995a, Tucker1998, Tucker2000, Tucker2001, Balcerzak2003, Balcerzak2004a, Balcerzak2009, Balcerzak2009a, Szalowski2012, Szalowski2014, Balcerzak2014b, Szalowski2014a,Dominguez2018}. 
For instance, the PA method has been first applied to the Heisenberg \cite{Strieb1963, Morita1966, Morita1972, Barry1975, Kinel1976, Balcerzak2009} and classical Ising \cite{Tucker1987, Morita1989, Katsura1996, Balcerzak2003} models. The studies included also dilute systems, for instance, involving the Ising model with long-range interactions \cite{Szalowski2014a}, as well as the Heisenberg anisotropic model \cite{Balcerzak2009a,Szalowski2011}. 
The Ising-type models with higher spins have also been considered \cite{Tucker1987}. Among other things, the Blume-Capel model with RKKY interaction \cite{Balcerzak2004a} or with mixed-spins \cite{Tucker2001}, as well as the Blume-Emery-Griffiths model \cite{Tucker1998, Tucker2000} have been investigated. Some papers have been devoted to thin films \cite{Balcerzak1995a}, especially to the bilayer \cite{Tucker1998, Balcerzak2009, Balcerzak2014b} and multilayer \cite{Szalowski2012,Szalowski2014} systems. 
Regarding quantum models, apart from the pure Heisenberg \cite{Balcerzak2009, Szalowski2012}, or mixed Heisenberg-Ising systems \cite{Balcerzak2014b}, the XXZ model with spin $S=1/2$ \cite{Bukman1991} and the transverse Ising model with a random field \cite{Ma1990} have been studied, together with XY model applied to describe a particular magnetocaloric material \cite{Konieczny2017}. Recently, the anisotropic Heisenberg model with spin $S=1$ has been investigated within this method \cite{Boubekri2017} and also a mixed-spin system with spins $1/2$ and 1 was considered \cite{Boubekri2019}. 
It is worth noticing that also some non-magnetic applications of PA have been developed, for example, for modelling of brain-computer interfaces \cite{Maren2016}.

In the course of these studies it turned out that PA is a very useful approach, giving correct physical results \cite{Morita1989}, at the cost of a moderate amount of effort. For example, it correctly predicts a lack of magnetism in 1D Ising model. Its results are also in agreement with Mermin-Wagner theorem for the isotropic Heisenberg model \cite{Mermin1966a}, since the PA predicts absence of magnetic ordering when the coordination number $z$ is less then $z=5$ \cite{Balcerzak2009}. It has also been shown that for the Blume-Emery-Griffiths model with $z=5$ the PA method gives exact results on a Bethe lattice \cite{Tucker1998}.
Moreover, regarding dilute systems, the PA method yields reasonable values of the critical concentrations \cite{Balcerzak2009a, Szalowski2014a}. An essential advantage of the PA method is that the Gibbs energy can be obtained in an analytical form, from which all thermodynamic properties can be calculated in a self-consistent way. The knowledge of the Gibbs energy is especially important for the 1st order phase transitions. Another merit is a possibility of calculation of spin-spin correlations, which are indispensable for investigations of such properties as magnetocaloric effect \cite{Szalowski2011,Szalowski2014, Boubekri2017}.

The shortcomings of the PA approach should also be mentioned. As an approximate method, it still overestimates the phase transition temperatures in comparison with higher-order cluster approximations \cite{Kikuchi1951}, as well as with Monte Carlo (MC) simulations \cite{Diaz2017,Diaz2018,Diaz2020}. The critical exponents calculated within this method are classical, i.e., the same as in MFA. Moreover, the PA is difficult to apply for the geometrically frustrated systems, where it should be adopted with a special care \cite{Balcerzak2014c}. Being aware of these shortcomings, the usefulness of the method is, to some extent, limited. Nevertheless, the advantage of the PA method over very common MFA is generally difficult to question.

A motivation behind further development of the PA method, and the present work, follows from the fact that, as far as we know, the Heisenberg model with spin $S=1$ and interaction of arbitrary range has not been studied by analytical approach. On the other hand, the models with spin $S>1/2$, with interactions extending beyond the first coordination zone,
including single-ion anisotropy and potentially also spin-space interaction anisotropies leading to XXZ model, should be useful for analytical description of modern ferromagnetic semiconductors of monolayer thickness. Such anisotropies are vital for emergence of the ferromagnetic ordering in 2D materials. At present, the search for monolayer magnetic materials by means of extensive DFT calculations covering a wide range of chemical compositions is reported \cite{Torelli2019,Guo2020,Zhang2019,Haastrup2018}, together with prediction of numerous novel structures and pronounced interest in maximizing the Curie temperature.

Taking these factors into account, the aim of the present paper is extension of the PA method for the quantum Heisenberg model with spin $S=1$. By contrast with the work of Boubekri \emph{at al.} \cite{Boubekri2017}, where the anisotropic Heisenberg model with only nearest-neighbour (NN) interactions was studied, we consider the isotropic model but with the interactions of arbitrary range, extending up to the arbitrary coordination zone. Moreover, the molecular field acting on the pair-cluster has two components in the present approach, which is appropriate for the case of spin value $S=1$. Namely, it consists of the ordinary bilinear field, and the quadrupolar field. We take note of the fact that the quadrupolar field was neglected in the paper \cite{Boubekri2017}, therefore it can be treated as a simplified version of our full and systematic method which we intend to present here. Our variational approach exploits the maximal number of independent quantum mechanical operators based on spin operator which can be utilized to construct the trial density matrix. The single-ion anisotropy term will also be taken into account in the present work, as well as the external magnetic field.

As an illustration of the developed method, we apply it for studies of the novel 2D magnetic semiconductor, CrIAs, described recently by Zhang et al. \cite{Zhang2019}. 
This ferromagnetic material with spin $S=1$ has been predicted theoretically by means of DFT technique, and seems to be very promising because of its high Curie temperature. The direct exchange and superexchange integrals parametrizing the magnetic interactions in this, so far hypothetical, 2D material have been determined in \cite{Zhang2019} by means of DFT calculations up to the third coordination zone, and they show strongly non-monotoniv behaviour with the distance. The single-ion anisotropy has also been determined there. Some MC studies performed in Ref.~\cite{Zhang2019} for CrIAs, using a finite supercell for the isotropic Heisenberg model with spin $S=1$, are presented only for the Curie temperature, magnetization and susceptibility. Therefore, we think that a more complete study of the magnetic properties of this material would be highly desirable, both due to the potential interest in the material itself and as an opportunity to illustrate our extension of the PA method on the example of a magnetic system with magnetic interactions not restricted to nearest-neighbours. The approach which we present seems to be a useful tool for studies of such type of ferromagnetic systems, aimed at complete description of their thermodynamics without resorting to simulational approaches. 

The paper is organized as follows: In the next, theoretical, section (\ref{theory}) the method is presented in detail. The extensive description of the formalism is intentional, since it may be helpful for the readers and potential users of the method. A part of derivations, concerning diagonalization of the pair Hamiltonian, has been moved to  Appendix~\ref{appendix}, and the formulas for thermodynamic properties are collected in Appendix~\ref{appendix B}. The third, numerical, section (\ref{num}) is devoted to application of the PA method for numerical calculations of the magnetic properties of CrIAs. The results of calculations of various thermodynamic properties  are illustrated in figures and discussed. In the final section (\ref{final}), the results of the paper are summarized and conclusions are drawn.

%section 2
\section{\label{theory}Theoretical model}

The present section contains a general derivation of the PA for the case of spin-1 Heisenberg model with interactions of arbitrary range. 

\subsection{The Pair Approximation method. General formulation}

We consider the quantum Heisenberg model with spin $S=1$, including the single-ion anisotropy term and the exchange interaction extending up to $n$-th coordination zone. The Hamiltonian is assumed in the form of:
\begin{equation}
{\mathcal H}=-\sum_{k=1}^{n}J_k\sum_{<i,j\in k>}^{Nz_k/2}\vec{S}_i \vec{S}_{j}  -A\sum_{i}^{N}\left(S_i^z\right)^2-h\sum_{i}^{N} S_i^z
\label{eq1}
\end{equation}
where $J_k$ ($k=1, ...,n$) is the exchange integral for the $k$-th coordination zone.  $A$-constant corresponds to the single-ion anisotropy and $h=-g\mu_{\rm B}H^z$ introduces the external magnetic field $H^z$. $N$ is the total number of spins in the system, whereas $z_k$ is the coordination number for the $k$-th zone (number of $k$-th NN).
$S_i^z$ denotes the $z$-component of the quantum spin $S=1$ in $i$-th lattice site, and takes the values of $S_i^z = \pm 1,0$).

The crucial theoretical problem is the Gibbs energy derivation. The Gibbs thermodynamic potential can be found on the basis of the formula:
\begin{equation}
G=\left<{\mathcal H}\right>-\sigma T, 
\label{eq2}
\end{equation}
where $\left<{\mathcal H}\right>$ (enthalpy) is the thermodynamic mean value of the Hamiltonian (\ref{eq1}) and $\sigma$ is the total entropy of the system. The entropy can be evaluated by the cumulant technique using cluster entropies \cite{Katsura1996}. Namely, in approximation where only the second order cumulants are taken into account we can write:
\begin{equation}
\sigma=N \sigma^{(1)} + \frac{N}{2}\sum_{k=1}^{n}z_k\left( \sigma_k^{(2)}-2 \sigma^{(1)}\right). 
\label{eq3}
\end{equation}
In (\ref{eq3}) $\sigma^{(1)}$ is the single-site entropy, which is not site dependent since we consider the ideal crystal. On the other hand, $\sigma_k^{(2)}$ ($k=1, ...,n$) are the pair entropies, where the pairs $<i,j\in k>$ are formed from the central spin, $\vec{S}_i$, and the other spin, $\vec{S}_{j}$, situated on the $k$-th coordination zone of the central spin.

Thus, the Gibbs energy (\ref{eq2}) can be presented as:
\begin{eqnarray}
G= -\frac{N}{2}\sum_{k=1}^{n} z_k J_k \left<\vec{S}_i \vec{S}_{j\in k}\right>^{(2)} -N A \left< \left(S_i^z\right)^2\right>^{(1)}
-N h \left< S_i^z\right>^{(1)} \nonumber\\
-N\left[ \frac{1}{2}\sum_{k=1}^{n} z_k \sigma_k^{(2)} +\left( 1-\sum_{k=1}^{n} z_k \right)\sigma^{(1)} \right]T.
\label{eq4}
\end{eqnarray}
In Eq.(\ref{eq4}) the local magnetization, $\left< S_i^z\right>^{(1)}$, the quadrupolar moment, $\left< \left(S_i^z\right)^2\right>^{(1)}$, and the single-site entropies $\sigma^{(1)}$ are calculated with the single-site density matrix, whereas the pair correlations, $ \left<\vec{S}_i \vec{S}_{j\in_k}\right>^{(2)}$, and the pair entropies, $\sigma_k^{(2)}$ ($k=1, ...,n$), should be calculated with the pair density matrices. These matrices are defined in the subsections below.

\subsection{Single-site density matrix}

The single-site density matrix for $i$-th lattice site is of the form:
\begin{equation}
\rho_i =\exp \left[ \beta \left( G^{(1)} - {\mathcal H}_{i} \right) \right] \;\;\;\;\;\; \left(\beta = 1/k_{\rm B}T  \right),
\label{eq5}
\end{equation}
where $G^{(1)}$ is the single-site Gibbs potential and ${\mathcal H}_{i}$ is the single-site trial Hamiltonian, which can be written as:
\begin{equation}
{\mathcal H}_{i}= - S_i^z \left(\lambda +h \right) - \left(S_i^z\right)^2 \left(\mu +A \right).
\label{eq6}
\end{equation}
The $\lambda$ parameter in Eq.(\ref{eq6}) corresponds to the linear molecular field and can be decomposed as:
\begin{equation}
\lambda = \sum_{k=1}^{n} z_k \lambda_k,
\label{eq7}
\end{equation}
where $\lambda_k$ ($k=1, ...,n$) describe contributions from a spin located on $k$-th coordination zone. Analogously, the $\mu$-parameter in the trial Hamiltonian corresponds to the quadrupolar component of the molecular field, and can be presented as:
\begin{equation}
\mu = \sum_{k=1}^{n} z_k \mu_k,
\label{eq8}
\end{equation}
where $\mu_k$ ($k=1, ...,n$) describe contributions to this field from a spin located on $k$-th coordination zone.\\
One has to emphasize here, that for an arbitrary spin $S$, in total $2S$ types of independent molecular fields can be included into the trial Hamiltonian. These fields are coupled to spin operators of the type $\left(S_i^z\right)^p$, where $p=1, ...,2S$. This comes from the fact that for spin $S$, the operators $\left(S_i^z\right)^p$ constitute independent operators for $p=1,\dots,2S$ \cite{Landau1977}. The number of these fields ($2S$), which are treated as variational parameters for the Gibbs energy, should be the same as the number of spin moments $\left\langle\left(S_i^z\right)^p\right\rangle$, for which $2S$ equations can be derived. Thus, in our case of $S=1$, two components of the molecular fields, i.e., $\lambda$ and $\mu$ in Eq.~(\ref{eq6}), form a complete representation. It should also be mentioned that in the usual MFA only one component of the molecular field (denoted by $\lambda$) is taken into account, independently on the spin value $S$. However, utilization of maximal number of variational parameters is more proper, since it exploits fully the number of independent quantum mechanical operators which can be used to construct the most general trial density matrix; moreover it makes the system more stable, as the Gibbs energy can be then lowered.

The single-site Gibbs energy, $G^{(1)}$, can be found from normalization condition for the density matrix (\ref{eq5}):
${\rm Tr}_i \, \rho_i =1.$
This condition leads to the formula:
\begin{equation}
G^{(1)}=- k_{\rm B}T \ln \left[{\rm Tr}_i\, e^{- \beta {\mathcal H}_{i}} \right] =- k_{\rm B}T \ln Z^{(1)},
\label{eq10}
\end{equation}
where the single-site statistical sum is:
\begin{equation}
Z^{(1)}={\rm Tr}_i\, e^{- \beta {\mathcal H}_{i}} = 2 e^{\beta \left( \mu+A \right)} \cosh \left[\beta \left( \lambda+h \right)  \right] +1.
\label{eq11}
\end{equation}
The single site entropy, $\sigma^{(1)}$, can  be found from the formula:
\begin{equation}
\sigma^{(1)}= -k_{\rm B} {\rm Tr}_i \left(\rho _i  \ln \rho _i \right)=-\frac{1}{T} G^{(1)} +\frac{1}{T} \left< {\mathcal H}_{i}\right>^{(1)},
\label{eq12}
\end{equation}
where $\left< {\mathcal H}_{i}\right>^{(1)}$ is the mean value of the trial Hamiltonian (\ref{eq6}) calculated with the help of the density matrix (\ref{eq5}), namely:
\begin{equation}
\left< {\mathcal H}_{i}\right>^{(1)}= -\left< S_i^z \right>^{(1)} \left(\lambda +h \right) - \left< \left(S_i^z\right)^2 \right>^{(1)} \left(\mu +A \right).
\label{eq14}
\end{equation}
The magnetization, $\left< S_i^z \right>^{(1)}$, and quadrupolar moment, $\left< \left(S_i^z\right)^2 \right>^{(1)}$, calculated with the matrix $\rho _i$ are of the form:
\begin{equation}
\left< S_i^z \right>^{(1)} \equiv m = \frac{2 \sinh \left[\beta \left( \lambda+h \right) \right]}{2 \cosh \left[\beta \left( \lambda+h \right) \right] + e^{-\beta \left( \mu+A \right)}},
\label{eq15}
\end{equation} 
and
\begin{equation}
\left< \left(S_i^z\right)^2 \right>^{(1)} \equiv q = \frac{2 \cosh \left[\beta \left( \lambda+h \right) \right]}{2 \cosh \left[\beta \left( \lambda+h \right) \right] + e^{-\beta \left( \mu+A \right)}},
\label{eq16}
\end{equation}
respectively.

\subsection{Pair density matrices}

In order to calculate the spin-pair correlations, $ \left<\vec{S}_i \vec{S}_{j\in k}\right>^{(2)}$, and the pair entropies, $\sigma_k^{(2)}$ ($k=1, ...,n$), in Eq.(\ref{eq4}) we introduce the pair density matrices as follows:
\begin{equation}
\rho_{i,j\in k} =\exp \left[ \beta \left( G_k^{(2)} - {\mathcal H}_{i,j\in k} \right) \right] \;\;\;\;\;\; \left(\beta = 1/k_{\rm B}T  \right),
\label{eq17}
\end{equation}
where $G_k^{(2)}$ is the two-site Gibbs potential corresponding to the trial Hamiltonian, ${\mathcal H}_{i,j \in k}$, of the ($i,j\in k$)-pair. The pair Hamiltonian is of the form:
\begin{equation}
{\mathcal H}_{i,j\in k} = -J_k \vec{S}_i \vec{S}_{j\in k} -\left(S_i^z + S_{j\in k}^z \right)\left(\lambda -\lambda_k +h \right) - \left[ \left(S_i^z\right)^2+ \left(S_{j\in k}^z\right)^2 \right] \left(\mu -\mu_k+A \right).
\label{eq18}
\end{equation}
The linear molecular fields, $\lambda -\lambda_k $, and quadrupolar fields, $\mu -\mu_k $, are acting on both sides of the  ($i,j\in k$)-pair, whereas the exchange interaction $J_k$ inside the pair is taken exactly.

From the normalization condition for the pair density matrix, ${\rm Tr}_{i,j\in k} \, \rho_{i,j\in k} =1$,
the pair Gibbs energy,  $G_k^{(2)}$, can be found in the form of:
\begin{equation}
G_k^{(2)}=- k_{\rm B}T \ln \left[{\rm Tr}_{i,j\in k}\, e^{- \beta {\mathcal H}_{i,j\in k}} \right] =- k_{\rm B}T \ln Z_k^{(2)}.
\label{eq20}
\end{equation}
However, in order to calculate the two-site statistical sum, $Z_k^{(2)}$, we must first diagonalize the pair Hamiltonian 
(\ref{eq18}). In this case the diagonalization procedure can be done analytically and its description is given in the Appendix~\ref{appendix}. Finally one obtains:
\begin{eqnarray}
Z_k^{(2)}&=&e^{\beta J_k\left( 2M_k+1\right)} \left[ 2 \cosh \left( 2 \beta J_k L_k \right) + e^{-2\beta J_k} \right] 
+4 e^{\beta J_k M_k} \cosh \left( \beta J_k \right) \cosh \left( \beta J_k L_k \right)\nonumber \\
&+&2 e^{\beta J_k \left(M_k-1/2 \right)} \cosh \left( \beta J_k \sqrt{\left( 1/2-M_k\right)^2 +2} \right),
\label{eq21}
\end{eqnarray}
for $k=1, ...,n$, where 
\begin{equation}
J_k L_k=  \lambda -\lambda_k +h,
\label{eq22}
\end{equation}
and
\begin{equation}
J_k M_k=  \mu -\mu_k +A 
\label{eq23}
\end{equation}
($\lambda$ and $\mu$ are given by Eqs.(\ref{eq7}) and (\ref{eq8}), respectively).

With the help of the pair density matrix (\ref{eq17}) the pair entropies, $\sigma_k^{(2)}$, can be found from the formula:
\begin{equation}
\sigma_k^{(2)}= -k_{\rm B} {\rm Tr}_{i,j\in k} \left(\rho _{i,j\in k}  \ln \rho _{i,j\in k} \right)
=-\frac{1}{T} G_k^{(2)} +\frac{1}{T} \left< {\mathcal H}_{i,j\in k}\right>^{(2)},
\label{eq24}
\end{equation}
where $\left< {\mathcal H}_{i,j\in k}\right>^{(2)}$ is the mean value of the trial pair Hamiltonian (\ref{eq18}) calculated with the pair density matrix $\rho _{i,j\in k}$, namely:
\begin{eqnarray}
\left< {\mathcal H}_{i,j\in k}\right>^{(2)}&=&-J_k \left< \vec{S}_i \vec{S}_{j\in k}\right>^{(2)} -\left(\left<S_i^z\right>^{(2)} +\left< S_{j\in k}^z\right>^{(2)} \right)\left(\lambda -\lambda_k +h \right) \nonumber \\
&-& \left[ \left<\left(S_i^z\right)^2\right>^{(2)}+ \left<\left(S_{j\in k}^z\right)^2\right>^{(2)} \right] \left(\mu -\mu_k+A \right).
\label{eq26}
\end{eqnarray}

The local magnetization, $\left<S_i^z\right>^{(2)}$, and the quadrupolar moment, $\left<\left(S_{i}^z\right)^2\right>^{(2)}$, can be easily calculated with the help of $\rho _{i,j\in k}$ and they are presented in the Appendix~\ref{appendix}.
Moreover, the spin-pair correlations can be found as a sum of longitudinal and perpendicular components:
\begin{equation}
\left< \vec{S}_i \vec{S}_{j\in k}\right>^{(2)} = \left< S_i^z S_{j\in k}^z\right>^{(2)} +\left< S_i^x S_{j\in k}^x
+ S_i^y S_{j\in k}^y\right>^{(2)}.
\label{eq29}
\end{equation}
The explicit forms of these correlations have been also presented in the Appendix~\ref{appendix}.  Thus, the pair entropies (\ref{eq24}) have been fully determined  by the pair density matrix  $\rho_{i,j\in k}$.

\subsection{Variational equations and the Gibbs energy in equilibrium}

Now, substituting the entropies given by (\ref{eq24}) and (\ref{eq12}) into expression (\ref{eq4}) we obtain:
\begin{eqnarray}
\frac{G}{N} = &-&\frac{1}{2}\sum_{k=1}^{n} z_k J_k \left<\vec{S}_i \vec{S}_{j\in k}\right>^{(2)} -A \left< \left(S_i^z\right)^2\right>^{(1)}- h \left< S_i^z\right>^{(1)}\nonumber \\
&-& \frac{1}{2}\sum_{k=1}^{n} z_k \left( G_k^{(2)} - \left< {\mathcal H}_{i,j\in k}\right>^{(2)}\right) +
\left( 1-\sum_{k=1}^{n} z_k \right)\left(G^{(1)} -\left< {\mathcal H}_{i}\right>^{(1)} \right) .
\label{eq30}
\end{eqnarray}
We see that the total Gibbs energy becomes a function of the molecular field parameters $\lambda_k$ and $\mu_k$ ($k=1, ...,n$).
Treating these parameters as variational variables with respect to which the total Gibbs energy should be minimized, we require satisfaction of $2n$ variational equations:
\begin{equation}
\frac{\partial G}{\partial \lambda_k} = 0,
\label{eq31}
\end{equation} 
and
\begin{equation}
\frac{\partial G}{\partial \mu_k} = 0   
\label{eq32}
\end{equation} 
for $k = 1, \ldots , n $.

Performing differentiation of Eq.(\ref{eq30}) it can be shown that the extreme conditions (\ref{eq31}) and (\ref{eq32}) are equivalent to the following equations:
\begin{equation}
m  =  m_k^{(2)}, 
\label{eq33}
\end{equation} 
and
\begin{equation}
q  =  q_k^{(2)}, 
\label{eq34}
\end{equation} 
for $k = 1, \ldots , n$. In Eqs.(\ref{eq33}) and (\ref{eq34}) $m$ and $q$ are given by (\ref{eq15}) and (\ref{eq16}), whereas $m_k^{(2)}$ and $q_k^{(2)}$ are given by (\ref{eq27}) and (\ref{eq28}), respectively. Eqs.(\ref{eq33}) and (\ref{eq34}) express the fact that the local magnetization and quadrupolar moment can be calculated either from the single-site or from the pair density matrix, giving the same result for any coordination zone $k$. This is, in fact, consistent with the pair density matrices property which, after partial reduction, should be equivalent to the single-site density matrix \cite{Morita1972}, namely: ${\rm Tr}_{j\in k} \, \rho_{i,j\in k} = \rho_{i}$.

Eqs.(\ref{eq33}) and (\ref{eq34}) for $k = 1, \ldots , n$ are treated as a set of $2n$ coupled equations from which the variational parameters $\lambda_k$ and $\mu_k$ can be numerically determined. With the help of these equations the total Gibbs energy per spin (\ref{eq30}) can be found in equilibrium in the final form:
\begin{equation}
\frac{G}{N} = \frac{1}{2}\sum_{k=1}^{n} z_k G_k^{(2)} + \left( 1-\sum_{k=1}^{n} z_k \right)G^{(1)},   
\label{eq36}
\end{equation} 
where $G^{(1)}$ is given by Eqs.(\ref{eq10}) and (\ref{eq11}), and $G_k^{(2)}$ for $k = 1, \ldots , n$ are given by Eqs.(\ref{eq20}) and (\ref{eq21}), respectively. Now, the energies $G^{(1)}$ and $G_k^{(2)}$ are fully known since  $\lambda_k$ and $\mu_k$ have been determined. In this method the total Gibbs energy  (\ref{eq36}) remains only a function of magnetic field $h$ and temperature $T$ and the  interaction parameters $J_k$ and $A$ of the Hamiltonian. The crystal structure is taken into account by the coordination numbers $z_k$.

\subsection{The phase transition (Curie) temperature}

In case of continuous phase transition, in the vicinity of phase transition temperature (when $h=0$, $T \to T_{\rm C}$, and $T < T_{\rm C}$) the molecular fields vanish, $\lambda_k \to 0$. Then, Eqs.(\ref{eq33}) for $k = 1, \ldots , n $ can be linearized with respect to $\lambda_k$ and we get:
\begin{equation}
\frac{\lambda}{2 + e^{-\beta_{\rm C} \left( \mu+A \right)}}= \frac{J_kL_k}{Z_k^{(2)}(0)}\left[2e^{\beta_{\rm C} J_k\left( 2M_k+1\right)}  + e^{\beta_{\rm C} J_k M_k} \cosh \left( \beta_{\rm C} J_k \right)  \right]
\label{eq46}
\end{equation}
where $\beta_{\rm C}=1/k_{\rm B} T_{\rm C}$, and $Z_k^{(2)}(0)$ is the statistical sum (\ref{eq21}) in which all $L_k$-parameters are set to zero:
\begin{eqnarray}
Z_k^{(2)}(0) &=& e^{\beta_{\rm C} J_k\left( 2M_k+1\right)} \left[2 + e^{-2\beta_{\rm C} J_k} \right] +4 e^{\beta_{\rm C} J_kM_k}
\cosh \left( \beta_{\rm C} J_k \right) \nonumber\\
&+&2e^{\beta_{\rm C} J_k\left( M_k-1/2\right)}\cosh \left[ \beta_{\rm C} J_k \sqrt{\left(1/2-M_k \right)^2 +2} \; \right]
\label{eq47}
\end{eqnarray}
Eq.(\ref{eq46}) is equivalent to the set of $n$ homogeneous linear equations for $\lambda_k$ of the form:
\begin{equation}
P_k \sum_{l=1}^{k-1}z_l \lambda_l + \left[\left(z_k-1\right) P_k+1 \right] \lambda_k +P_k \sum_{l=k+1}^{n}z_l \lambda_l =0,
\label{eq48}
\end{equation}
for $k = 1, \ldots , n $, where:
\begin{equation}
P_k = 1- \frac{2 + e^{-\beta_{\rm C} \left( \mu+A \right)}}{Z_k^{(2)}(0)}\left[ 2e^{\beta_{\rm C} J_k\left( 2M_k+1\right)}
+ e^{\beta_{\rm C} J_k M_k} \cosh \left( \beta_{\rm C} J_k \right)\right].
\label{eq49}
\end{equation}
The homogeneous set of linear equations (\ref{eq48}) can be formally written as:
\begin{equation}
{\mathcal M} \cdot \vec{\Lambda}=0,
\label{eq50}
\end{equation}
where
\begin{equation}
\vec{\Lambda}=\left[ \begin{array}{c}
\lambda_1\\
\vdots \\
\lambda_n\\
\end{array}
 \right],
\label{eq51}
\end{equation}
and  matrix ${\mathcal M}$ has the elements:
\begin{eqnarray}
{\mathcal M}_{k,k^{\prime}}&=&P_k z_{k^{\prime}} \;\;\;\;\;\;\;\;({\rm for} \;\;k \ne k^{\prime}); \nonumber\\
{\mathcal M}_{k,k}&=& \left(z_k-1\right) P_k+1 .
\label{eq52}
\end{eqnarray}
In order to solve (\ref{eq48}) the determinant of matrix ${\mathcal M}$ is set to zero:
\begin{equation}
\det {\mathcal M} =0,
\label{eq53}
\end{equation}
which is valid for arbitrary number of coordination zones $n$. For example, for $n=3$, from  (\ref{eq53}) one obtains:
\begin{eqnarray}
\left(z_1+z_2+z_3-1 \right)P_1P_2P_3-\left(z_1+z_2-1 \right)P_1P_2 - \left(z_1+z_3-1 \right)P_1P_3 
- \left(z_2+z_3-1 \right)P_2P_3 \nonumber\\
+\left(z_1-1 \right)P_1+\left(z_2-1 \right)P_2 +\left(z_3-1 \right)P_3 +1 =0, \;\;\;\;\
\label{eq54}
\end{eqnarray}
where $P_k$ are given by Eq.(\ref{eq49}). Eq.(\ref{eq53}) allows determination of the Curie temperature provided that $\mu_k$-parameters are known. The $n$ supplementary equations for $\mu_k$ are obtained from (\ref{eq34}) where, for $T_{\rm C}$, we put $\lambda_k \to 0$. Then we get the formulas:
\begin{eqnarray}
\frac{2}{2 + e^{-\beta_{\rm C} \left( \mu+A \right)}} &=&
\frac{1}{Z_k^{(2)}(0)} \Bigg\{ e^{\beta_{\rm C} J_k\left( 2M_k+1\right)} \left[ 2 + e^{-2\beta_{\rm C} J_k} \right] 
+2 e^{\beta_{\rm C} J_k M_k} \cosh \left( \beta_{\rm C} J_k \right) \nonumber \\
&+& e^{\beta_{\rm C} J_k \left(M_k-1/2 \right)} \bigg[\cosh \left( \beta_{\rm C} J_k \sqrt{\left( 1/2-M_k\right)^2 +2} \right)\nonumber \\
&+& \frac{2 M_k-1}{2\sqrt{\left( 1/2-M_k\right)^2 +2} } 
\sinh \left( \beta_{\rm C} J_k \sqrt{\left( 1/2-M_k\right)^2 +2} \right)\bigg] \Bigg\} 
\label{eq55}
\end{eqnarray}
for $k = 1, \ldots , n $, where $M_k= \left( \sum_{l=1}^{n} z_l \mu_l -\mu_k +A \right)/J_k$ from Eq.(\ref{eq23}). Thus, 
Eqs.(\ref{eq55}) and (\ref{eq53}) form a complete set of $n+1$ equations for the Curie temperature determination.

%section 3
\section{\label{num}Numerical results and discussion}

\begin{figure}[ht]
\begin{center}
\includegraphics[width=0.8\textwidth]{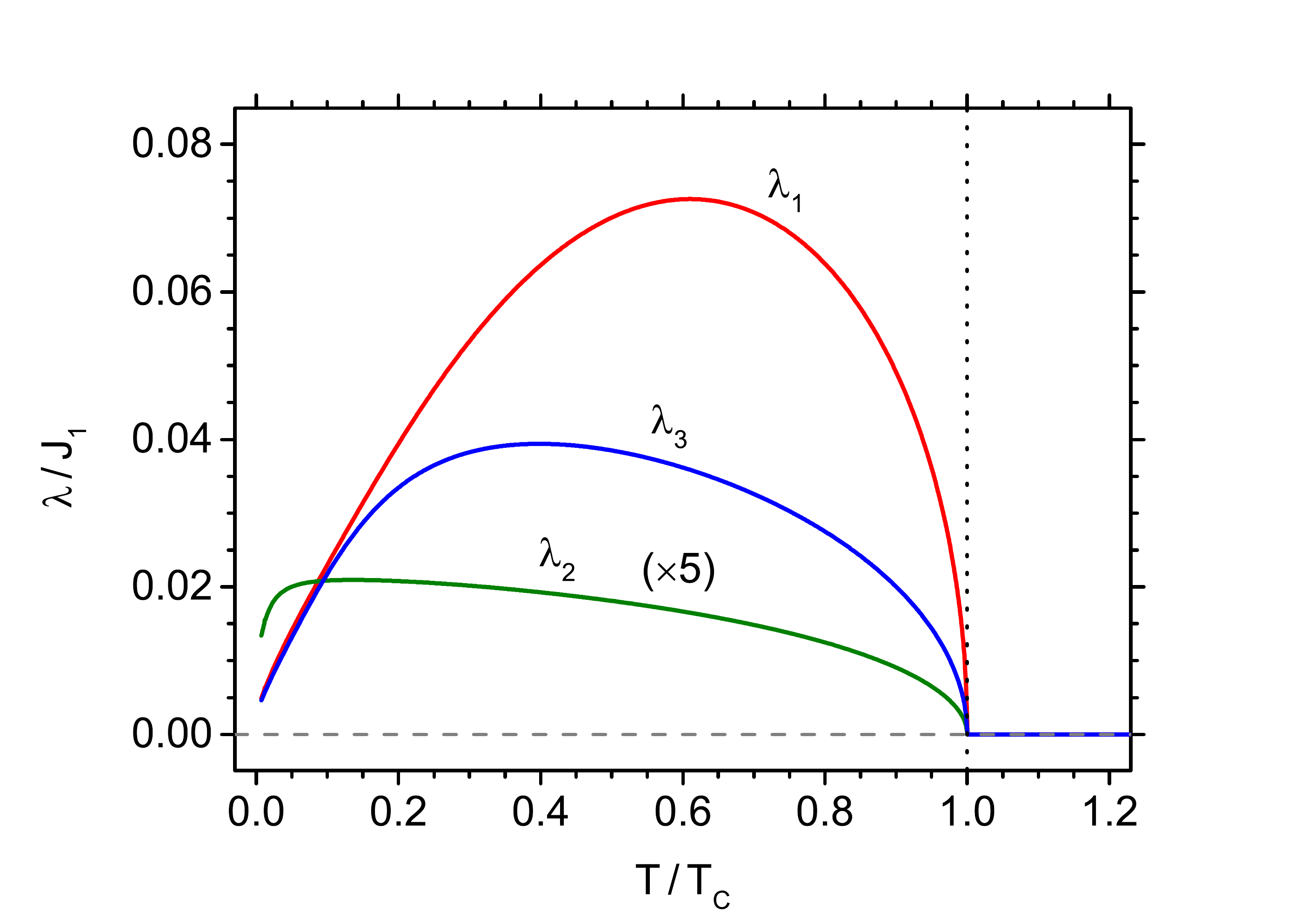}
\caption{\label{fig1}The dependence of the normalized variational parameter $\lambda$ for 1NN, 2NN and 3NN on the normalized temperature. Note that the data for 2NN were multiplied by 5 for clarity of presentation. The critical (Curie) temperature is marked with vertical dashed line.}
\end{center}
\end{figure}

\begin{figure}[ht]
\begin{center}
\includegraphics[width=0.8\textwidth]{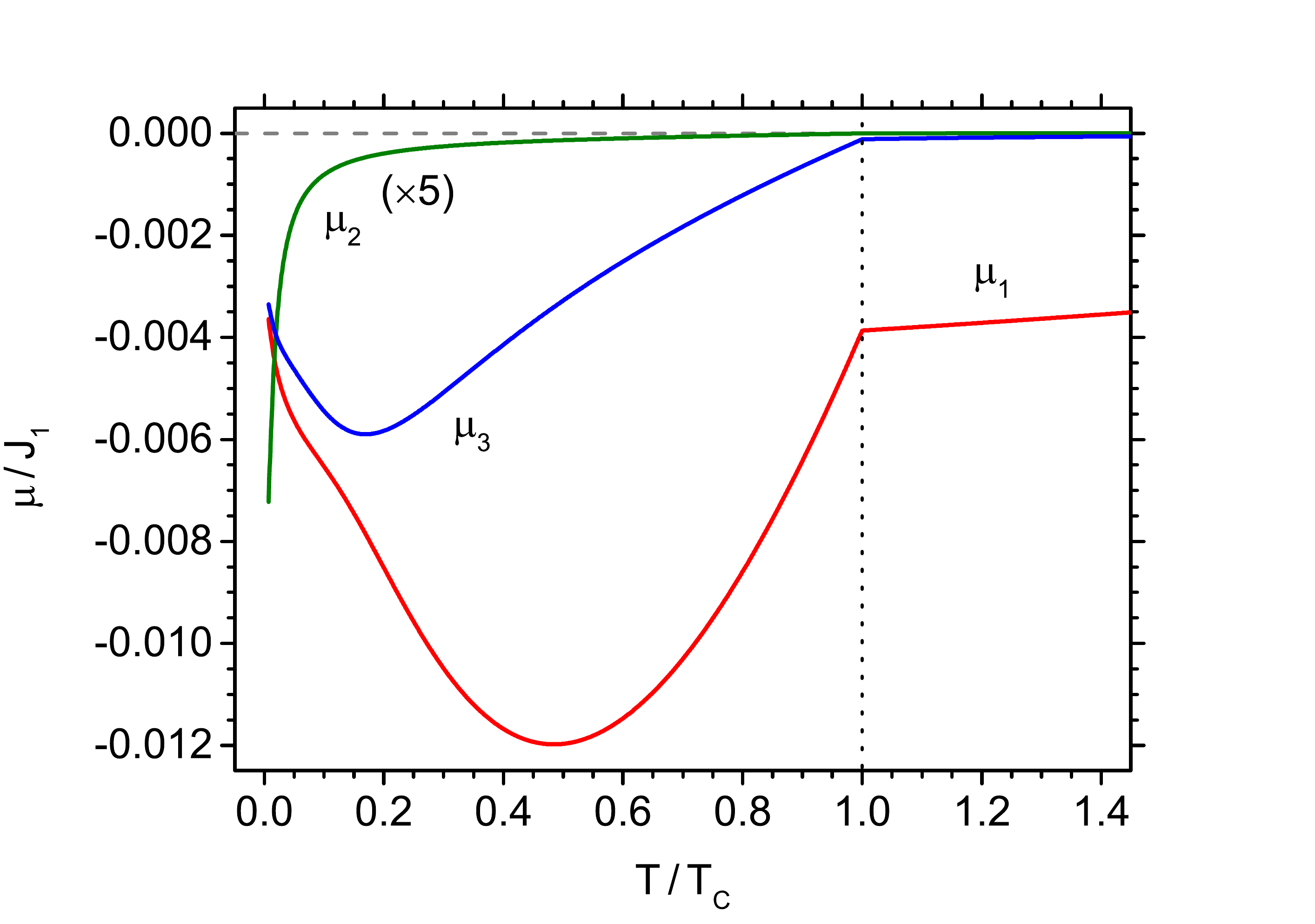}
\caption{\label{fig2}The dependence of the normalized variational parameter $\mu$ for 1NN, 2NN and 3NN on the normalized temperature. Note that the data for 2NN were multiplied by 5 for clarity of presentation. The critical (Curie) temperature is marked with vertical dashed line.}
\end{center}
\end{figure}

The numerical calculations are performed for the 2D magnetic semiconductor CrIAs \cite{Zhang2019} to illustrate the developed formalism of PA method. According to DFT results, the exchange integrals are equal to $J_1$=289.6 meV, $J_2$=1.3 meV, and $J_3$=17.6 meV, whereas the single-ion anisotropy was found as $A$=6.1 meV. The same quantities, but normalized to NN exchange integral $J_1$, have the dimensionless values: $J_1/J_1=1$, $J_2/J_1=4.489\cdot 10^{-3}$, and $J_3/J_1=6.0773\cdot 10^{-2}$, for the  1st (1NN), 2nd (2NN), and 3rd (3NN) coordination zone, respectively, whereas the reduced single-ion anisotropy parameter is $A/J_1=2.1064\cdot 10^{-2}$. The above dimensionless parameters are the only input data necessary (and fixed) for computations based on the formalism presented in previous section. The 2D crystalline lattice of CrIAs is characterized by the coordination numbers $z_1=2$, $z_2=4$, and $z_3=2$. The calculations are performed vs. dimensionless temperature $k_{\rm B}T/J_1$, whereas the normalized external field is $h/J_1$.

All the temperature dependences of the thermodynamic parameters are presented as a function of the normalized temperature $T/T_{C}$, where $T_{C}$ is the critical (Curie) temperature of the continuous phase transition between the ferromagnetic state and the paramagnetic state. Its reduced value calculated from Eq.(\ref{eq54}) equals to $k_{\rm B}T_{C}/J_1$=0.24783. In physical units this would be approximately 833 K. It is about 27\% larger than the value 655 K, or $k_{\rm B}T_{C}/J_1$=0.19490 in dimensionless units, predicted in \cite{Zhang2019} by MC simulation. Let us note here that some overestimation of $T_{C}$-value in comparison with MC results is a common feature of approximate methods. For instance, the MFA, which is the most simple method, yields the following expression for the Curie temperature in the present model: 
$k_{\rm B}T_{C}\left[ 2+\exp \left(-A/k_{\rm B}T_{C} \right) \right]=2\left(z_1J_1+z_2J_2+z_3J_3 \right)$. Its solution gives 
$k_{\rm B}T_{C}/J_1$=1.43331, which is even several times greater than our PA result.

Let us commence the discussion from the variational parameters, $\lambda$ and $\mu$, the determination of which is a key point in calculation of the Gibbs energy and all the thermodynamic quantities of interest. The parameter $\lambda$ couples to the spin in a similar way as the usual molecular field and is a dominant parameter in PA formulation for arbitrary spin. Fig.~\ref{fig1} presents the dependence of individual parameters $\lambda_k$ for $k=1,2,3$ originating from 1NN, 2NN and 3NN, respectively. It can be observed that at the critical (Curie) temperature $T_{C}$ all the parameters vanish continuously and remain zero for higher temperatures. For $T<T_{C}$ they exhibit a non-monotonic behaviour, as they reach the local maxima. The magnitude of each parameter $\lambda_k$ is correlated with the magnitude of the exchange integral $J_k$ - the largest $\lambda_k$ corresponds to the strongest $J_k$. Moreover, a local maximum is reached at highest temperature for the parameter $\lambda$ corresponding to strongest magnetic exchange integral. When the temperature decreases, the parameters decrease linearly, reaching the non-zero values at $T=0$. The overall behaviour is similar to the variability of analogous parameter for PA applied to the case of spin $S=1/2$ (see Ref.~\cite{Balcerzak2009}).

The second variational parameter, $\mu$, couples to the square of the spin and has the sense of quadrupolar molecular field. Its variability with the temperature can be followed in Fig.~\ref{fig2}, for all three parameters $\mu_k$ corresponding to 1NN, 2NN and 3NN, respectively. Similarly to the case of $\lambda_k$, the most pronounced magnitudes of $\mu$ correspond to the strongest exchange integral $J_k$. In the whole range of $T$, the parameters take negative values. Unlike $\lambda$ parameters, at $T\geq T_{C}$ they do not vanish; instead they decrease asymptotically to zero. The magnitudes of $\mu_2$ and $\mu_3$ are greatly reduced at $T\geq T_{C}$, whereas $\mu_1$ remains significant at $T=T_{C}$ and very slowly tends to 0 when the temperature increases.

\begin{figure}[ht]
\begin{center}
\includegraphics[width=0.8\textwidth]{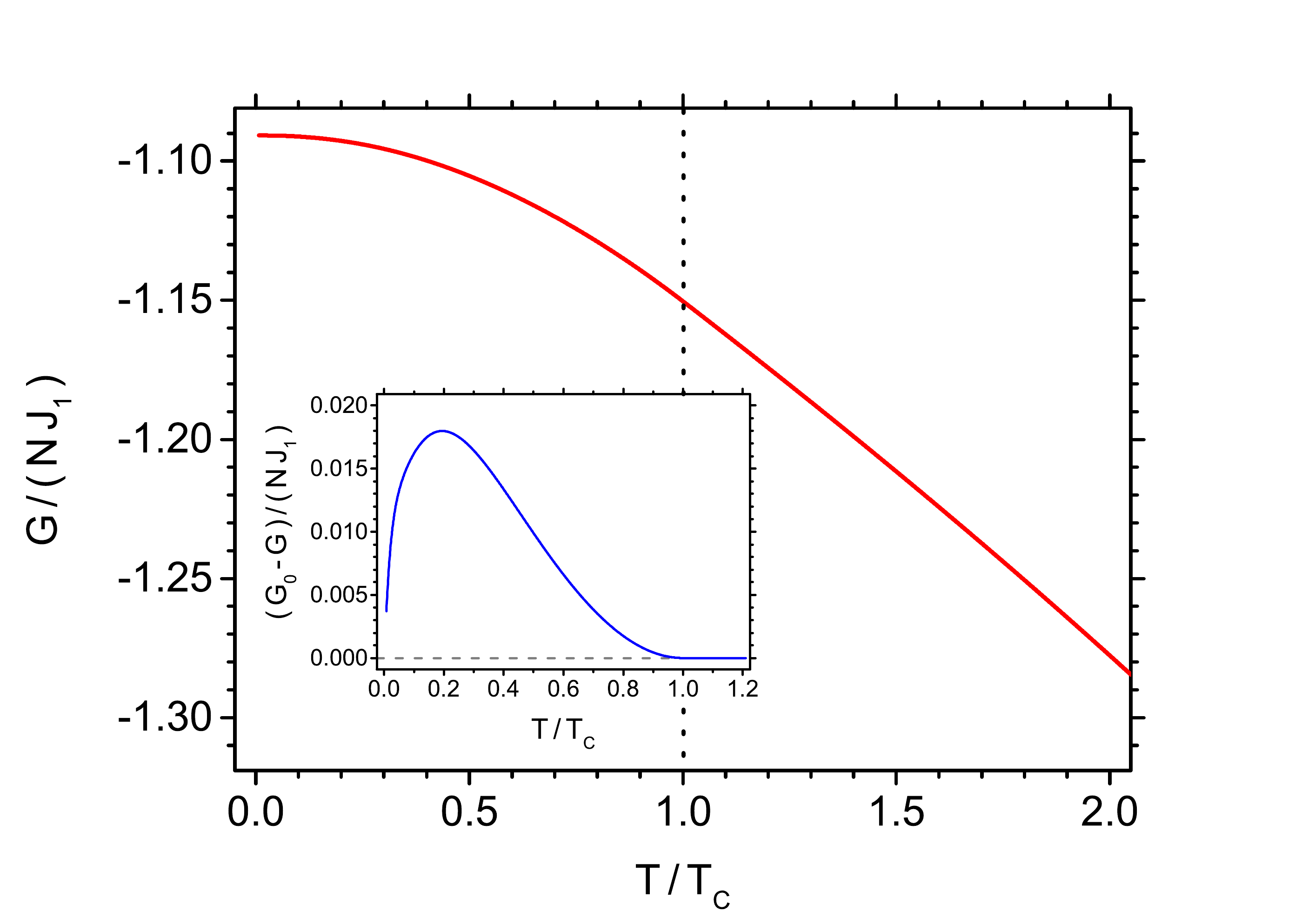}
\caption{\label{fig3}The dependence of the normalized Gibbs free energy per site on the normalized temperature. The inset shows the normalized difference between the Gibbs free energy of paramagnetic phase and ferromagnetic phase. The critical (Curie) temperature is marked with vertical dashed line.}
\end{center}
\end{figure}

\begin{figure}[ht]
\begin{center}
\includegraphics[width=0.8\textwidth]{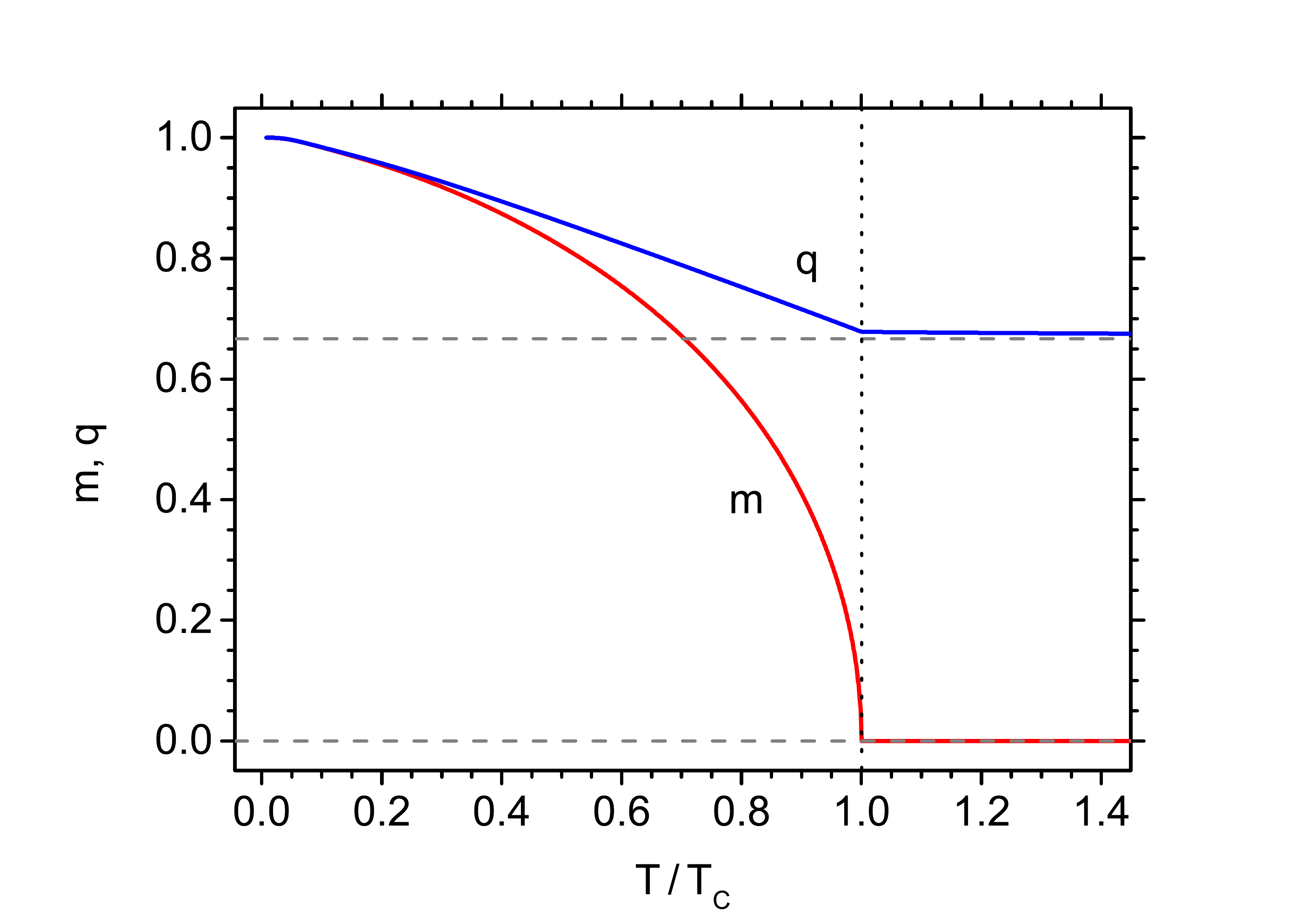}
\caption{\label{fig4}The dependence of magnetization and quadrupolar moment on the normalized temperature. The critical (Curie) temperature is marked with vertical dashed line.}
\end{center}
\end{figure}

The fundamental thermodynamic potential in the present study is the Gibbs free energy, the variational minimization of which leads to the equations determining the parameters $\lambda$ and $\mu$. The ability to calculate the Gibbs energy within PA enables the complete characterization of the thermodynamics of the system. The Gibbs energy is shown as a function of the temperature in Fig.~\ref{fig3}. It is a monotonic, decreasing function of the temperature, with the slope tending to zero at $T=0$ (thus corresponding to zero entropy at $T=0$). It is instructive to compare the Gibbs energy $G$ of the physically stable phase (plotted in the main panel of Fig.~\ref{fig3}) with the Gibbs energy for paramagnetic phase $G_0$ (the phase characterized with $\lambda=0$ and $\mu<0$). The difference $G-G_0$ is much smaller than the magnitude of $G$ itself, thus it is plotted separately, in the inset to Fig.~\ref{fig3}. It can be seen that for $T<T_{C}$ we obtain $G<G_0$, so that the ferromagnetic phase is thermodynamically stable for this range of temperatures. At $T=T_{C}$ the Gibbs energies of both ferromagnetic and paramagnetic phase coincide; it is also seen that the slope of $G_0-G$ takes the value of 0 at $T=T_{C}$, so that the slopes of both $G$ and $G_0$ are the same at $T=T_{C}$, as expected for second-order (continuous) phase transition. It might be mentioned that the knowledge of the Gibbs energy enables the search for discontinuous phase transitions, however, for the present parameters of the model such transitions were not noticed. The calculation of the Gibbs energy enables the systematic study of all other thermodynamic quantities.

The temperature behaviour of the parameters $\lambda$ and $\mu$ is directly reflected in the temperature dependence of the magnetization and quadrupolar moment. Fig.~\ref{fig4} depicts the variation of both quantities with the temperature. At $T=0$ both $m$ and $q$ take the saturated values of 1, corresponding to the saturated ferromagnetic ordering. The magnetization decreases continuously to 0, reaching this value at $T_{C}$ (reflecting the dominant influence of $\lambda$ on $m$). Above $T_{C}$ the paramagnetic phase with $m=0$ is stable. This is in a contrast with MC calculations for finite clusters \cite{Zhang2019} where the magnetization at $T_{C}$ still had remarkable value. The quadrupolar moment also decreases with the temperature, but slower than magnetization; at $T_{C}$ it takes the value of 0.6783 (slightly above $2/3$) and its slope changes. For $T>T_{C}$, $q$ slowly tends asymptotically to $2/3$ (shown with the horizontal dashed line in Fig.~\ref{fig4}), reflecting the important effect of $\mu$ on $q$ when $\lambda$ equals to 0 in paramagnetic phase.

\begin{figure}[ht]
\begin{center}
\includegraphics[width=0.99\textwidth]{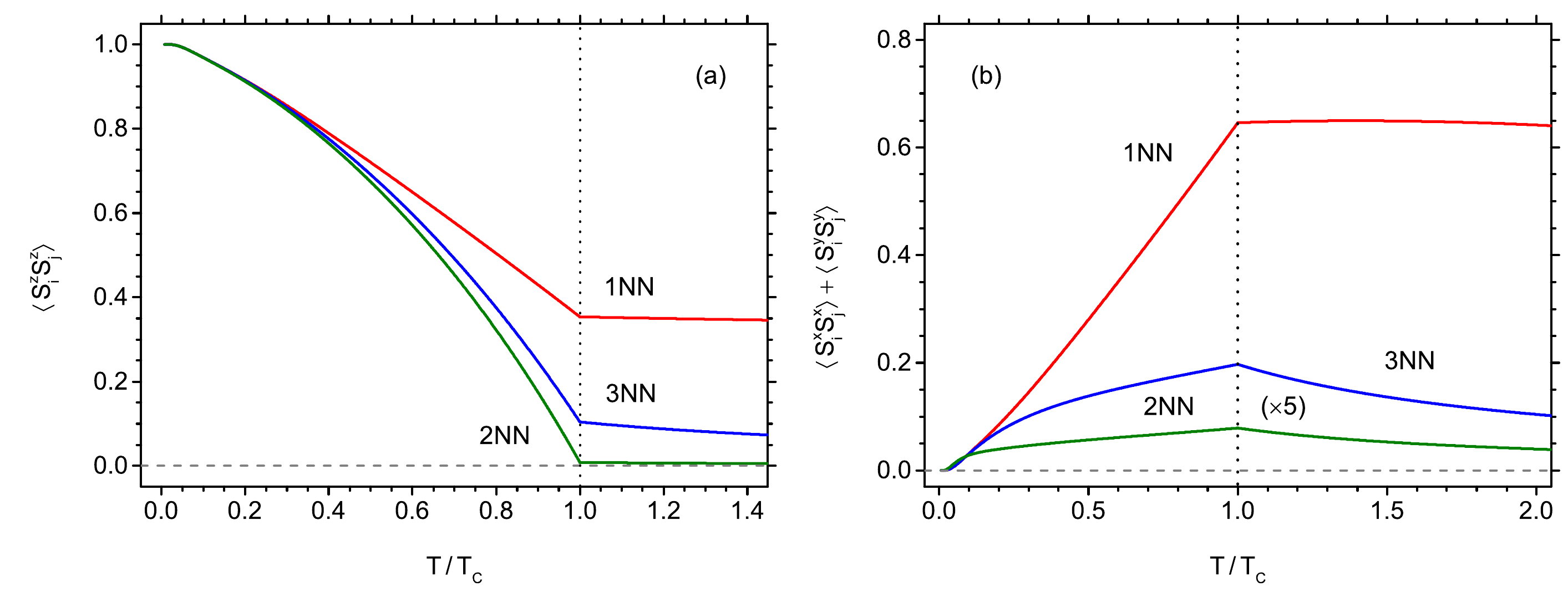}
\caption{\label{fig5}The dependence of the longitudinal (a) and transverse (b) spin-spin correlations for 1NN, 2NN and 3NN on the normalized temperature. Note that the data for transverse correlations for 2NN were multiplied by 5 for clarity of presentation. The critical (Curie) temperature is marked with vertical dashed line.}
\end{center}
\end{figure}

The behaviour of the spin-spin correlations is shown in Fig.~\ref{fig5}. The longitudinal correlations (between $z$ components of spins) for 1NN, 2NN and 3NN are presented in Fig.~\ref{fig5}(a). At $T=0$ they reach the saturated value of 1 and decrease with increasing temperature. The strength of the correlations between $k$-th NN reflects the strength of the exchange integral $J_k$. At the critical temperature the correlations do not drop down to 0, and for $T>T_{C}$ their values tend asymptotically, but very slowly, to 0 (note that the 1NN correlations remain significant at $T_{C}$ and still amount there to 0.35357). The behaviour of the perpendicular correlations (between $x$ components or between $y$ components of spins) is markedly different, as it is visible in Fig.~\ref{fig5}(b). Namely, at $T=0$ these correlations are equal to 0. The fact that all the correlations in the ground state are between $z$ components of spins proves that the ferromagnetically ordered state is of Ising type (due to the presence of the non-zero crystal field $A$). Going up to the critical temperature the perpendicular correlations increase and take the maximal values at $T_{C}$ (note that for 1NN the corresponding magnitude of correlation is 0.64603, so that almost twice more than the value of longitudinal correlations at the same temperature). For $T>T_{C}$, the perpendicular correlations slowly decrease, finally tending asymptotically to 0 (but especially the magnitude for 1NN remains considerably large up to high temperatures).

\begin{figure}[ht]
\begin{center}
\includegraphics[width=0.8\textwidth]{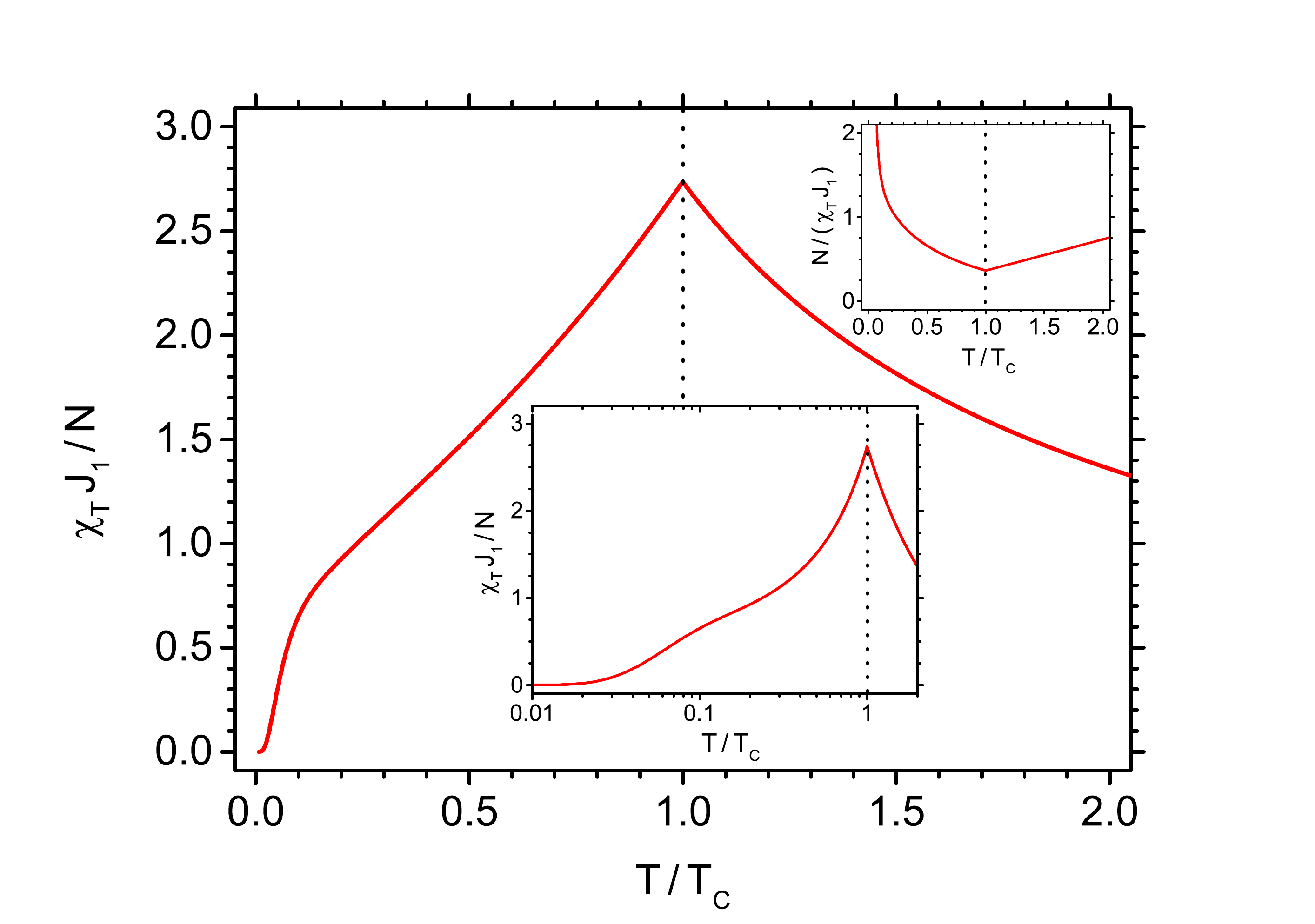}
\caption{\label{fig6}The dependence of the normalized isothermal magnetic susceptibility on the normalized temperature. Central inset shows the dependence with logarithmic scale for the temperature. Right inset presents the temperature dependence of the inverse of susceptibility. The critical (Curie) temperature is marked with vertical dashed line.}
\end{center}
\end{figure}

One of the response functions of common interest for magnetic systems is the isothermal magnetic susceptibility, which is plotted in Fig.~\ref{fig6}. It is evident that this quantity takes a broad maximum at the critical temperature. The low-temperature behaviour can be traced in details in the inset showing the same data in logarithmic temperature scale to emphasize the low-temperature range. Within this range, a weak maximum-like feature is noticeable at $T/T_{C}\simeq $0.1. On the other hand, the behaviour of the susceptibility for $T>T_{C}$ is well illustrated in the second inset, where the inverse of susceptibility is plotted as a function of the temperature. It is evident that above $T_{C}$, the susceptibility follows the Curie-Weiss law, since $1/\chi_{T}$ is a linear function of $T$. It is also worth noticing that the decrease of susceptibility vs. temperature in paramagnetic state is slower in our case than that predicted in Ref.~\cite{Zhang2019}. 

\begin{figure}[ht]
\begin{center}
\includegraphics[width=0.8\textwidth]{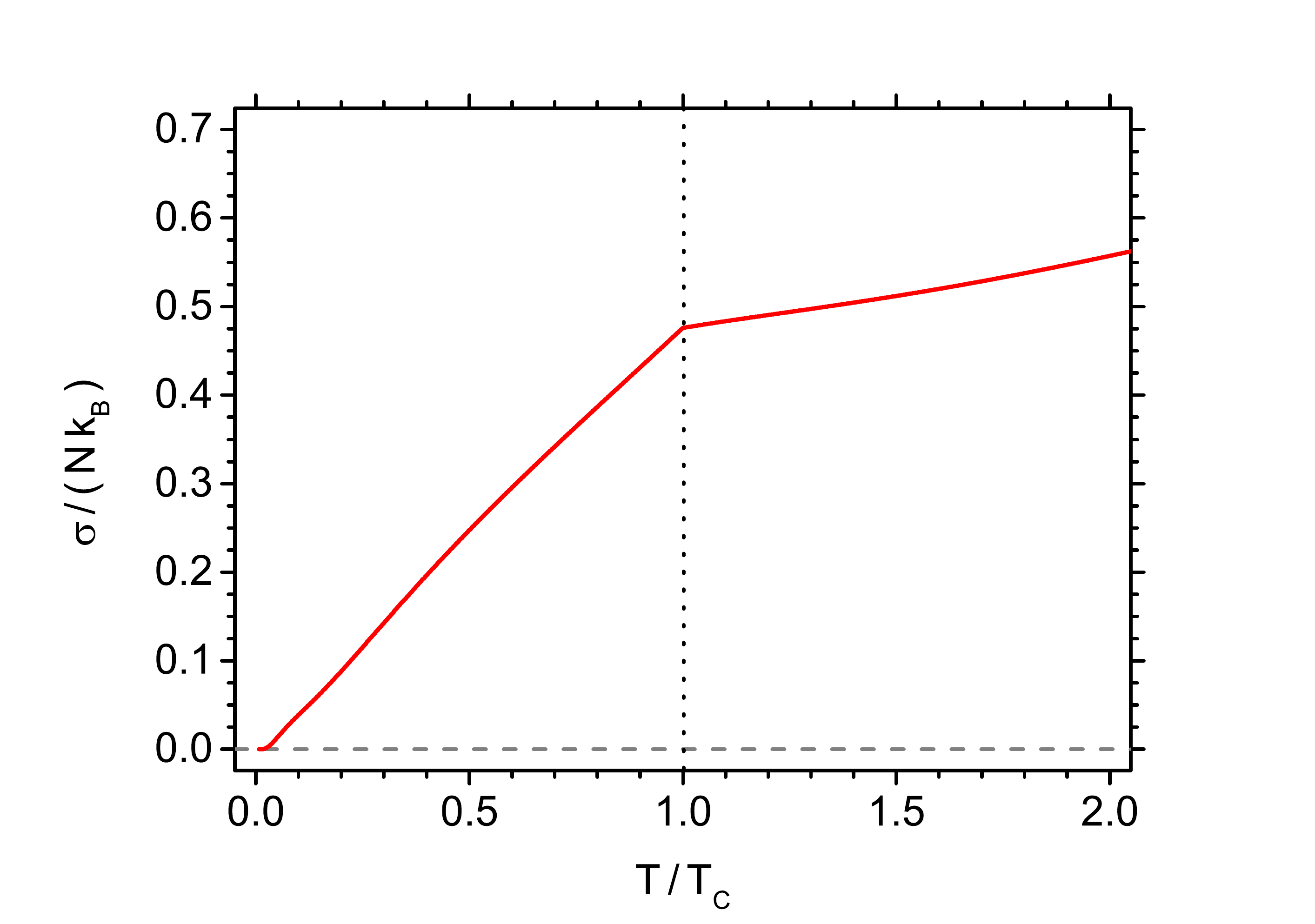}
\caption{\label{fig7}The dependence of the normalized entropy per site on the normalized temperature. The critical (Curie) temperature is marked with vertical dashed line.}
\end{center}
\end{figure}

\begin{figure}[ht]
\begin{center}
\includegraphics[width=0.8\textwidth]{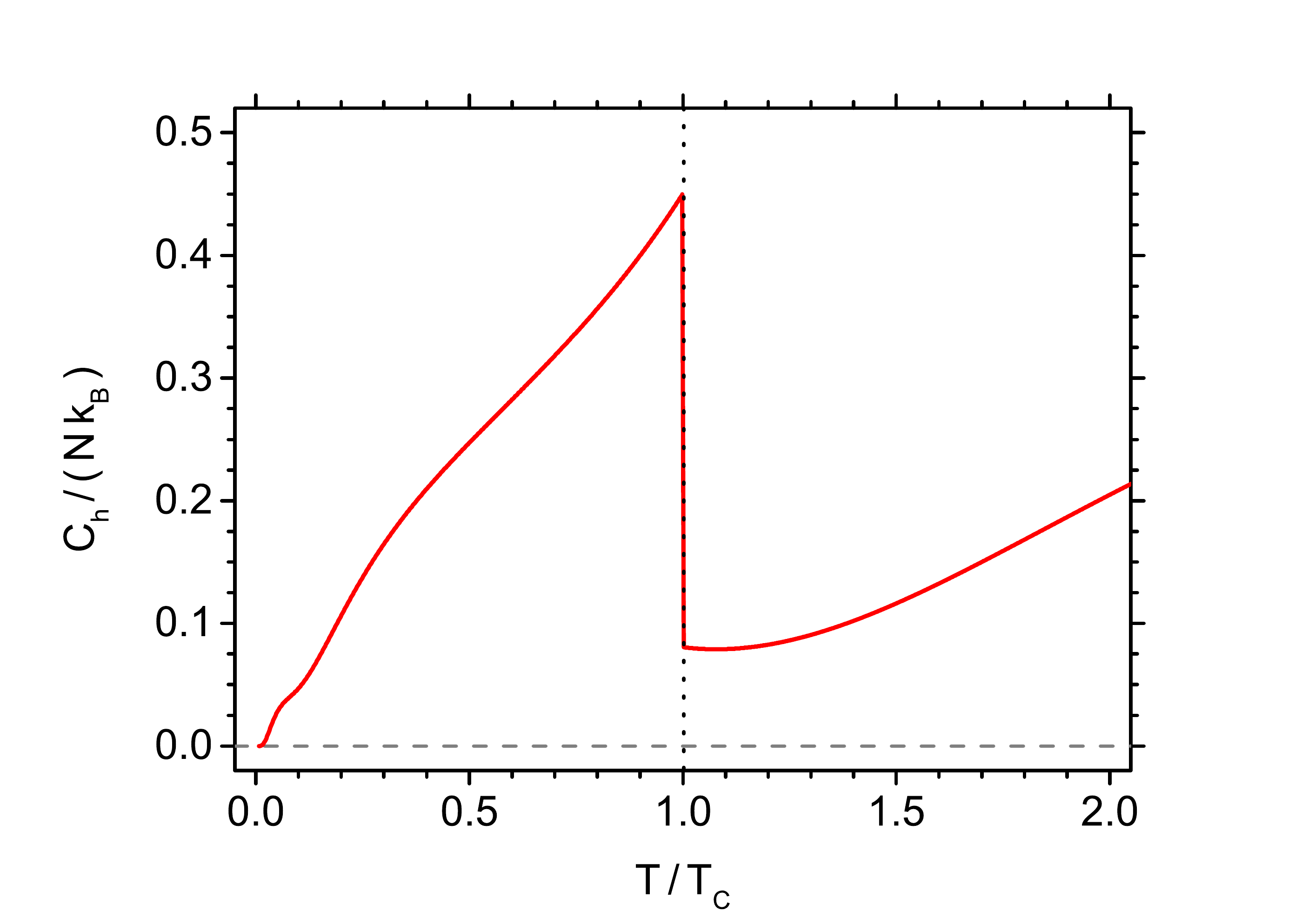}
\caption{\label{fig8}The dependence of the normalized specific heat on the normalized temperature. The critical (Curie) temperature is marked with vertical dashed line.}
\end{center}
\end{figure}

The behaviour of the entropy as a function of the temperature is plotted in Fig.~\ref{fig7}. For the temperatures below $T_{C}$, a quasi-linear increase is observed, with a trace of a low temperature feature mentioned above. At critical temperature the entropy is continuous, since we deal with the second order phase transition. The normalized entropy value at $T_{C}$ is 0.47589. It amounts to about 0.43318 of the saturation entropy, which reaches the value of $\ln 3\simeq$ 1.09861. Therefore, a very significant part of the total entropy (more than a half) comes from the paramagnetic phase range. For $T>T_{C}$, the entropy rises slowly, tending to the saturation value, and this dependency is correlated with analogous slow behaviour of other quantities in the paramagnetic regime.

The behaviour of the specific heat at constant magnetic field, $C_{h}$, is closely related to the behaviour of entropy and can be followed in Fig.~\ref{fig8}. Below the critical temperature, $C_{h}$ rises in linear-like manner, with a trace of a feature seen at low temperatures. For $T>T_{C}$, after reaching the discontinuity at $T_{C}$, the specific heat again increases (and a wide paramagnetic peak is present at the temperature as high as $T/T_{C}\simeq$ 4). However, this part of the curve is not presented here since it extends over an unphysical (too high) temperature range for the compound in question.  The noticeable values of the specific heat at the paramagnetic phase correlate well with the increase in entropy in this range of temperatures.

%section 4
\section{\label{final}Summary and conclusion}

In the paper the quantum Heisenberg model with spin $S=1$, including the single-ion anisotropy and arbitrary range of interaction, has been studied. The improved PA method has been developed, in which besides the bilinear molecular field, the quadrupolar field has been introduced. This improvement has enabled a self-consistent determination of the quadrupolar moment $q$ (see Eq.(\ref{eq34})), thus utilizing the full potential of the PA method in case of spin $S=1$. The theoretical method has been presented in detail; in particular, the Gibbs free-energy, as well as the formulas for all the basic thermodynamic quantities, have been derived. In Appendix~\ref{appendix}, the diagonalization procedure for the spin-pair clusters has been presented, which may be useful for studies of similar problems concerning the localized spin models.

In the Numerical Results and Discussion section (\ref{num}) the formalism has been applied for studies of the novel hypothetical ferromagnetic semiconductor CrIAs - a 2D system the existence of which has been predicted by DFT technique. All the thermodynamic quantities relevant to describe the magnetic properties for a system were calculated. It has been found that the Curie temperature obtained by the present method exceeds by about 27\% the value resulting from MC simulations for finite clusters \cite{Zhang2019}. As mentioned in the Introduction (\ref{intro}), when comparing the PA and MC methods, the phase transition temperatures calculated in the PA are somewhat overestimated \cite{Diaz2017,Diaz2018,Diaz2020}. However, the difference is only quantitative, and, as far as we know, all the results obtained in the PA are qualitatively correct. Therefore, the method can be recommended for the comprehensive studies of low-dimensional magnets, where some simpler analytical approaches, like MFA, are not appropriate. 

The numerical results obtained for the theoretical system CrIAs have elucidated the main magnetic properties of this material. The experimental verification of theoretical predictions will be possible when such a compound is synthesized in practice. We think that the presented method can be further developed and applied also for other real systems, where the Heisenberg model is applicable.

It should be emphasized that our method is particularly useful for description of the 2D magnetic systems modelled with Heisenberg Hamiltonian with arbitrary-range ferromagnetic interactions and with single-ion anisotropy. This anisotropy is a key factor allowing the emergence of ferromagnetic ordering in 2D systems, which constitute a highly interesting and rapidly developing class of modern nanomaterials \cite{Torelli2019,Guo2020,Zhang2019,Li2019e,Gong2019}. Therefore, the method would be potentially useful for obtaining a complete thermodynamic characterization of this class of prospective magnetic systems, supplementing the approaches strongly focused at sole calculation of the critical (Curie) temperature of monolayer materials \cite{Torelli2018,Olsen2019,Lu2019}.  

Since the Gibbs free-energy of the present model is available, the phase diagrams and the thermodynamic properties can be theoretically investigated, whereas the parameters of interactions are changed. For instance, the effect of the single-ion anisotropy on the phase diagrams can be studied, and possible first-order transitions are able to be determined \cite{Strecka}. However, the full discussion of the influence of the modified value of the anisotropy and variation of the long-range interaction parameters on the phase diagram of the model would exceed the scope of our present work. Therefore, in order to fully explore the model from a pure theoretical point of view, a separate work would be necessary. Such exploration would be particularly fruitful after inclusion of the spin-space interaction anisotropy.

Regarding the further studies, the PA method can still be developed to describe the models with higher quantum spins. Such direction is well justified by the fact that numerous predicted 2D ferromagnetic materials indicate the presence of localized spins even higher than one \cite{Torelli2019,Guo2020}. Another direction might be connected with the description of the dilute magnets, for example related to more conventional diluted magnetic semiconductors. The inclusion of such parameters as directional anisotropies as well as biquadratic interactions is also possible.

\appendix
\section {\label{appendix}Diagonalization of the pair Hamiltonian}

The spin matrices for the quantum spin $S=1$ have the form:
\begin{equation}
S^x=\frac{1}{\sqrt{2}}\left( \begin{array}{ccc}
0 & 1 & 0\\
1 & 0 & 1\\
0 & 1 & 0\\
\end{array}
\right); \;\;\;\;\;
S^y=\frac{1}{\sqrt{2}}\left( \begin{array}{ccc}
0 & -i & 0\\
i & 0 & -i\\
0 & i & 0\\
\end{array}
\right); \;\;\;\;\;
S^z=\left( \begin{array}{ccc}
1 & 0 & 0\\
0 & 0 & 0\\
0 & 0 & -1\\
\end{array}
\right); 
\label{a1}
\end{equation}
and the identity matrix is:
\begin{equation}
I=\left( \begin{array}{ccc}
1 & 0 & 0\\
0 & 1 & 0\\
0 & 0 & 1\\
\end{array}
\right); 
\label{a2}
\end{equation}
In order to distinguish between the spins in $i$-th and $j\in k$-th site of the $(i,j\in k)$-pair we define the matrices:
\begin{equation}
S_{i}^{\alpha}=S^{\alpha}\otimes I, \;\;\;\;\;\;\; {\rm and} \;\;\;\;\;\;\;  S_{j\in k}^{\alpha}=I \otimes S^{\alpha}
\;\;\;\;\;\;\;(\alpha = x,y,z)
\label{a3}
\end{equation}
where $\otimes$ is the outer product. For the outer product and the ordinary product of matrices the following relationship holds:
\begin{equation}
\left(A \otimes B \right) \cdot \left(A^{\prime} \otimes B^{\prime} \right)=
\left(A \cdot A^{\prime} \right) \otimes \left(B \cdot B^{\prime} \right).
\label{a4}
\end{equation}
Using the formula (\ref{a4}) the ordinary product of the type $S_{i}^{\alpha} S_{j\in k}^{\alpha}$ can be presented as:
\begin{equation}
S_{i}^{\alpha} S_{j\in k}^{\alpha}= \left(S^{\alpha} \otimes I \right) \left(I \otimes S^{\alpha} \right)=
S^{\alpha} I \otimes I S^{\alpha}= S^{\alpha} \otimes S^{\alpha}.
\label{a5}
\end{equation}
By the same token
\begin{equation}
\left(S_{i}^{\alpha} \right)^2=
S_{i}^{\alpha} S_{i}^{\alpha}= \left(S^{\alpha} \right)^2 \otimes I,\;\;\;\;\;\;\; {\rm and} \;\;\;\;\;\;\;
\left(S_{j\in k}^{\alpha} \right)^2= I \otimes \left(S^{\alpha} \right)^2.
\label{a6}
\end{equation}
Using the relations (\ref{a3}), (\ref{a5}) and (\ref{a6}) the trial Hamiltonian of the $(i,j\in k)$-pair (Eq.(\ref{eq18})) can be presented as:
\begin{eqnarray}
{\mathcal H}_{i,j\in k}= -J_k \left(S^x \otimes S^x + S^y \otimes S^y + S^z \otimes S^z\right) -
\left(S^z \otimes I + I \otimes S^z \right)\left(\lambda -\lambda_k +h \right) \nonumber\\
- \left[ \left(S^z\right)^2 \otimes I+ I \otimes \left(S^z\right)^2 \right] \left(\mu -\mu_k+A \right).
\label{a7}
\end{eqnarray}

We see that the pair Hamiltonian presents a $9 \times 9$ matrix. Its explicit form is the following:
\begin{equation}
\frac{{\mathcal H}_{i,j\in k}}{J_k}=\left[ \begin{array}{ccccccccc}
d_1 & 0 & 0 & 0 & 0 & 0 & 0 & 0 & 0\\
0 & d_2 & 0 & -1 & 0 & 0 & 0 & 0 & 0\\ 
0 & 0 & d_3 & 0 & -1 & 0 & 0 & 0 & 0\\
0 & -1 & 0 & d_4 & 0 & 0 & 0 & 0 & 0\\
0 & 0 & -1 & 0 & d_5 & 0 & -1 & 0 & 0\\ 
0 & 0 & 0 & 0 & 0 & d_6 & 0 & -1 & 0\\
0 & 0 & 0 & 0 & -1 & 0 & d_7 & 0 & 0\\
0 & 0 & 0 & 0 & 0 & -1 & 0 & d_8 & 0\\ 
0 & 0 & 0 & 0 & 0 & 0 & 0 & 0 & d_9\\
\end{array}
\right], 
\label{a8}
\end{equation}

The diagonal elements can be listed as:
\begin{eqnarray}
d_1 &=& -2L_k -2M_k -1  \nonumber\\
d_2 &=& -L_k -M_k  \nonumber\\
d_3 &=& -2M_k +1  \nonumber\\
d_4 &=& -L_k -M_k   \nonumber\\
d_5 &=& 0  \nonumber\\
d_6 &=& L_k -M_k   \nonumber\\
d_7 &=& -2M_k +1  \nonumber\\
d_8 &=& L_k -M_k  \nonumber\\
d_9 &=& 2L_k -2M_k -1,
\label{a9}
\end{eqnarray}
where $L_k$ and $M_k$ are defined by Eqs.(\ref{eq22}) and (\ref{eq23}), respectively. 

Matrix (\ref{a8}) can be diagonalized analytically. As a result we obtain the following eigenvalues:
\begin{eqnarray}
E_1 &=&d_1 = -2L_k -2M_k -1  \nonumber\\
E_2 &=& -L_k -M_k -1  \nonumber\\
E_3 &=&d_3 = -2M_k +1  \nonumber\\
E_4 &=& -L_k -M_k +1  \nonumber\\
E_5 &=& 1/2 - M_k + \sqrt{\left( 1/2-M_k\right)^2 +2}   \nonumber\\
E_6 &=& L_k -M_k -1  \nonumber\\
E_7 &=& 1/2 - M_k - \sqrt{\left( 1/2-M_k\right)^2 +2} \nonumber\\
E_8 &=& L_k -M_k +1 \nonumber\\
E_9 &=&d_9 = 2L_k -2M_k -1.
\label{a10}
\end{eqnarray}

The statistical sum, $Z_k^{(2)}$, for the $(i,j\in k)$-pair can be calculated from the formula:
\begin{equation}
Z_k^{(2)}= \sum_{r=1}^{9} e^{- \beta J_k E_r}.
\label{a11}
\end{equation}
Substituting (\ref{a10}) into (\ref{a11}) we obtain $Z_k^{(2)}$ in the form of Eq.(\ref{eq21}).

The eigenvectors, $\mid \Psi_r \big >$,  for $r = 1, \ldots , 9 $, diagonalizing (\ref{a8}) and corresponding to the eigenvalues (\ref{a10}) have the explicit form:

\begin{equation}
\mid \Psi_1 \big > = \left[ \begin{array}{c}
1\\0\\0\\0\\0\\0\\0\\0\\0\\
\end{array}
\right]; \;\;\;\;\;\;\;  
\mid \Psi_2 \big > = \frac{1}{\sqrt{2}} \left[ \begin{array}{c}
0\\1\\0\\1\\0\\0\\0\\0\\0\\
\end{array}
\right]; \;\;\;\;\;\;\; 
\mid \Psi_3 \big > = \frac{1}{\sqrt{2}} \left[ \begin{array}{c}
0\\0\\1\\0\\0\\0\\-1\\0\\0\\
\end{array}
\right];
\nonumber
\label{a12}
\end{equation}

\begin{equation}
\mid \Psi_4 \big > = \frac{1}{\sqrt{2}} \left[ \begin{array}{c}
0\\1\\0\\-1\\0\\0\\0\\0\\0\\
\end{array}
\right]; \;\;\;\;\;\;\;  
\mid \Psi_5 \big > = \frac{1}{\sqrt{\left(E_5\right)^2/2+1}} \left[ \begin{array}{c}
0\\0\\-E_5/2\\0\\1\\0\\-E_5/2\\0\\0\\
\end{array}
\right]; \;\;\;\;\;\;\; 
\mid \Psi_6 \big > = \frac{1}{\sqrt{2}} \left[ \begin{array}{c}
0\\0\\0\\0\\0\\1\\0\\1\\0\\
\end{array}
\right];
\nonumber
\label{a12b}
\end{equation}

\begin{equation}
\mid \Psi_7 \big > = \frac{1}{\sqrt{\left(E_7\right)^2/2+1}} \left[ \begin{array}{c}
0\\0\\-E_7/2\\0\\1\\0\\-E_7/2\\0\\0\\
\end{array}
\right]; \;\;\;\;\;\;\; 
\mid \Psi_8 \big > = \frac{1}{\sqrt{2}} \left[ \begin{array}{c}
0\\0\\0\\0\\0\\1\\0\\-1\\0\\
\end{array}
\right]; \;\;\;\;\;\;\;  
\mid \Psi_9 \big > = \left[ \begin{array}{c}
0\\0\\0\\0\\0\\0\\0\\0\\1\\
\end{array}
\right].
\label{a12c}
\end{equation}

The above eigenvectors are normalized, orthogonal and form a complete set. With the help of them various statistical averages can be calculated. In general, for a given operator $\hat{O}$ its statistical average calculated with the pair density matrix is found from the expression:
\begin{equation}
\left< \hat{O} \right>^{(2)}  =\frac{1}{Z_k^{(2)}} \sum_{r=1}^{9} \big < \Psi_r \mid \hat{O}\mid \Psi_r \big >
e^{- \beta J_k E_r},
\label{a13}
\end{equation}
where the matrix elements $\big < \Psi_r \mid \hat{O}\mid \Psi_r \big >$ can be determined using (\ref{a12}). For instance, the pair magnetization per spin can be found as:
\begin{equation}
m_k^{(2)} = \frac{1}{2} \left <S_i^{z} + S_{j\in k}^{z} \right>^{(2)} =
\frac{1}{2Z_k^{(2)}} \sum_{r=1}^{9} \big < \Psi_r \mid S^{z}\otimes I + I \otimes S^{z}\mid \Psi_r \big >
e^{- \beta J_k E_r},
\label{a14}
\end{equation}
and the result is:

\begin{equation}
\left< S_i^z \right>^{(2)} \equiv m_k^{(2)} = \frac{2}{Z_k^{(2)}}\left[e^{\beta J_k\left( 2M_k+1\right)} \sinh \left( 2 \beta J_k L_k \right) + e^{\beta J_k M_k} \cosh \left( \beta J_k \right) \sinh \left( \beta J_k L_k \right) \right].
\label{eq27}
\end{equation}
By the same token, the quadrupolar moment of the pair per spin can be found as:
\begin{equation}
q_k^{(2)} = \frac{1}{2} \left <\left(S_i^{z}\right)^2 + \left(S_{j\in k}^{z}\right)^2 \right>^{(2)} =
\frac{1}{2Z_k^{(2)}} \sum_{r=1}^{9} \big < \Psi_r \mid \left(S^{z}\right)^2\otimes I + 
I \otimes \left(S^{z}\right)^2\mid \Psi_r \big >
e^{- \beta J_k E_r},
\label{a15}
\end{equation}
with the result: 
\begin{eqnarray}
\left<\left(S_{i}^z\right)^2\right>^{(2)}\equiv q_k^{(2)} &=& \frac{1}{Z_k^{(2)}} \Bigg\{ e^{\beta J_k\left( 2M_k+1\right)} \left[ 2 \cosh \left( 2 \beta J_k L_k \right) + e^{-2\beta J_k} \right] \nonumber \\
&+&2 e^{\beta J_k M_k} \cosh \left( \beta J_k \right) \cosh \left( \beta J_k L_k \right) \nonumber \\
&+& e^{\beta J_k \left(M_k-1/2 \right)} \bigg[\cosh \left( \beta J_k \sqrt{\left( 1/2-M_k\right)^2 +2} \right)\nonumber \\
&+& \frac{2 M_k-1}{2\sqrt{\left( 1/2-M_k\right)^2 +2} } 
\sinh \left( \beta J_k \sqrt{\left( 1/2-M_k\right)^2 +2} \right)\bigg] \Bigg\}. 
\label{eq28}
\end{eqnarray}
Regarding the spin-spin longitudinal correlations we can now write:
\begin{equation}
\left <S_i^{z}  S_{j\in k}^{z} \right>^{(2)} =
\frac{1}{Z_k^{(2)}} \sum_{r=1}^{9} \big < \Psi_r \mid S^{z}\otimes S^{z}\mid \Psi_r \big >
e^{- \beta J_k E_r}.
\label{a16}
\end{equation}
The final result is obtained in the form of:
\begin{eqnarray}
\left <S_i^{z}  S_{j\in k}^{z} \right>^{(2)}&=& \frac{1}{Z_k^{(2)}} \Bigg\{ e^{\beta J_k\left( 2M_k+1\right)} \left[ 2 \cosh \left( 2 \beta J_k L_k \right) - e^{-2\beta J_k} \right] \nonumber \\
&-& e^{\beta J_k \left(M_k-1/2 \right)} \bigg[\cosh \left( \beta J_k \sqrt{\left( 1/2-M_k\right)^2 +2} \right)\nonumber \\
&+& \frac{2 M_k-1}{2\sqrt{\left( 1/2-M_k\right)^2 +2} } 
\sinh \left( \beta J_k \sqrt{\left( 1/2-M_k\right)^2 +2} \right)\bigg] \Bigg\}. 
\label{a17}
\end{eqnarray}

Similarly, for perpendicular correlations we have:
\begin{equation}
\left <S_i^{x}  S_{j\in k}^{x}  + S_i^{y}  S_{j\in k}^{y}\right>^{(2)} =
\frac{1}{Z_k^{(2)}} \sum_{r=1}^{9} \big < \Psi_r \mid S^{x}\otimes S^{x} + S^{y}\otimes S^{y}\mid \Psi_r \big >
e^{- \beta J_k E_r},
\label{a18}
\end{equation}
with the final result:

\begin{eqnarray}
\left <S_i^{x}  S_{j\in k}^{x}  + S_i^{y}  S_{j\in k}^{y}\right>^{(2)} &=&
\frac{4}{Z_k^{(2)}} \Bigg\{ e^{\beta J_k M_k} \sinh \left( \beta J_k \right) \cosh \left(\beta J_k L_k \right) \nonumber \\
&+& e^{\beta J_k \left(M_k-1/2 \right)} 
\frac{1}{\sqrt{\left( 1/2-M_k\right)^2 +2} } 
\sinh \left( \beta J_k \sqrt{\left( 1/2-M_k\right)^2 +2} \right) \Bigg\}\nonumber\\.
\label{a19}
\end{eqnarray}

Summation of (\ref{a17}) and (\ref{a19}) gives the total spin-spin correlations $\left< \vec{S}_i \vec{S}_{j\in k}\right>^{(2)}$
for $k = 1, \ldots , n $, according to Eq.(\ref{eq29}).\\

\section {\label{appendix B} Calculation of thermodynamic properties}

All thermodynamic properties can be self-consistently derived from the Gibbs energy given by Eq.(\ref{eq36}).

\begin {bf} Magnetization: \end {bf}
Magnetization $M$ of the system can be found from the first derivative of the Gibbs energy with respect to the external field:
\begin{equation}
M = -\left(\frac{\partial G}{\partial h}\right)_T = Nm.
\label{eq37}
\end{equation}
The result is $Nm$, where the single-site magnetization $m$ is given either by the formula (\ref{eq15}) or, equivalently, by Eq.(\ref{eq27}). The equivalence of (\ref{eq15}) and (\ref{eq27}) is provided by the relation (\ref{eq33}).

\begin {bf} Susceptibility: \end {bf} 
Isothermal susceptibility $\chi_T$ can be found after second differentiation of the Gibbs energy with respect to $h$:
\begin{equation}
\chi_T = -\left(\frac{\partial^2 G}{\partial h^2}\right)_T = N \left(\frac{\partial m}{\partial h}\right)_T.
\label{eq38}
\end{equation}
The most simple way to compute this quantity is the numerical differentiation of magnetization.

\begin {bf} Entropy: \end {bf}
Entropy $\sigma$ can be simply found from the first derivative of the Gibbs energy with respect to temperature:
\begin{equation}
\sigma = -\left(\frac{\partial G}{\partial T}\right)_h.
\label{eq39}
\end{equation}
 From (\ref{eq36}) and (\ref{eq39}) we obtain:
\begin{equation}
\frac{\sigma}{N k_{\rm B}} = -\frac{1}{k_{\rm B} T} \left[\frac{1}{2}\sum_{k=1}^{n} \frac{z_k R_k^{(2)}}{Z_k^{(2)}} + 
\frac{ 1-\sum_{k=1}^{n} z_k }{Z^{(1)}}R^{(1)}  \right] - \frac{G}{N k_{\rm B} T},
\label{eq40}
\end{equation}
where the statistical sum $Z^{(1)}$ is given by (\ref{eq11}), and $Z_k^{(2)}$ for $k = 1, \ldots , n $ are of the form (\ref{eq21}). The coefficients $R^{(1)}$ and $R_k^{(2)}$ in Eq.(\ref{eq40}) are then given by:
\begin{equation}
R^{(1)}= 2 e^{\beta \left( \mu +A \right)} \left\{\cosh \left[\beta \left(\lambda +h\right) \right]\left(\mu +A\right) +
\sinh \left[\beta \left(\lambda +h\right) \right]\left(\lambda +h\right)  \right\},
\label{eq41}
\end{equation}
and
\begin{eqnarray}
R_k^{(2)}&=& e^{\beta J_k\left( 2M_k+1 \right)}\bigg\{ \left[ 2\cosh \left(2 \beta J_k L_k \right) + e^{-2\beta J_k}\right]
J_k\left( 2M_k+1 \right)+ 4\sinh \left(2 \beta J_k L_k \right)J_k L_k \nonumber \\
&-&2J_k e^{-2\beta J_k} \bigg\}\nonumber \\
&+& 4e^{\beta J_kM_k} \bigg[ \cosh \left(\beta J_k \right)\cosh \left(\beta J_k L_k \right)J_k M_k 
+ \sinh \left(\beta J_k \right)\cosh \left(\beta J_k L_k \right)J_k \nonumber \\
&+& \cosh \left(\beta J_k \right)\sinh \left(\beta J_k L_k \right)J_k L_k ) \bigg] \nonumber \\
&+& 2 e^{\beta J_k\left( M_k-1/2\right)} \bigg\{ \cosh \left[ \beta J_k \sqrt{\left(1/2-M_k \right)^2 +2} \; \right]
J_k \left(M_k-1/2 \right)\nonumber \\
&+& \sinh \left[ \beta J_k \sqrt{\left(1/2-M_k \right)^2 +2} \;\right]J_k\sqrt{\left(1/2-M_k \right)^2 +2}
\; \bigg\}
\label{eq42}
\end{eqnarray}
(for $k = 1, \ldots , n $).
In an equivalent way, entropy can be found from Eq.(\ref{eq3}), since the density matrices have been fully characterized.

\begin {bf} Enthalpy: \end {bf} Having calculated the Gibbs energy and entropy, the enthalpy $H$ can be easily found from Eq.(\ref{eq2}), namely:
\begin{equation}
H=\left<{\mathcal H}\right> =G+\sigma T. 
\label{eq43}
\end{equation}
Alternatively, the enthalpy could be calculated as a mean value of the Hamiltonian (\ref{eq1}), since all correlations 
$\left< \vec{S}_i \vec{S}_{j\in k}\right>^{(2)}$ are already known (Eq.(\ref{eq29})), as well as the magnetization, $\left< S_i^z \right>^{(1)}=m$, which is given by Eq.(\ref{eq15}) (or (\ref{eq27})), whereas the quadrupolar moment, $\left<\left(S_{i}^z\right)^2\right>^{(1)}=q$,  is given by Eq.(\ref{eq16}) (or (\ref{eq28})).

\begin {bf} The magnetic heat capacity: \end {bf} The heat capacity at constant field, $C_h$, is given by:
\begin{equation}
C_h = -T\left(\frac{\partial^2 G}{\partial T^2}\right)_h = T\left(\frac{\partial \sigma}{\partial T}\right)_h,
\label{eq44}
\end{equation}
and can be computed by numerical differentiation of the entropy $\sigma$.
Alternatively, the heat capacity could be found from the formula:
\begin{equation}
C_h = \left(\frac{\partial H}{\partial T}\right)_h,
\label{eq45}
\end{equation}
provided the enthalpy $H$ has been already calculated.

\end{document}